\documentclass[12pt]{article}

\usepackage{graphicx}
\usepackage{amsmath}
\usepackage{amssymb}

\def\CR{\nonumber \\}
\def\eq#1{(\ref{#1})}
\def\[#1\]{\begin{align}#1\end{align}}
\def\hM{{\hat M}}
\def\hP{{\hat P}}
\def\hH{{\hat {\cal H}}}
\def\hJ{{\hat {\cal J}}}
\def\hD{{\hat {\cal D}}}
\def\hqH{{\hat H}}
\def\hqJ{{\hat J}}

\begin{document}
\begin{titlepage}
\title{
\hfill\parbox{4cm}{ \normalsize YITP-13-36}\\
\vspace{1cm} Quantum canonical tensor model\\ and an exact wave function}
\author{Naoki {\sc Sasakura}
\thanks{\tt sasakura@yukawa.kyoto-u.ac.jp}
\\[15pt]
{\it Yukawa Institute for Theoretical Physics, Kyoto University,}\\
{\it Kyoto 606-8502, Japan}}
\date{}
\maketitle
\thispagestyle{empty}
\begin{abstract}
\normalsize
Tensor models in various forms are being studied as models of quantum gravity.
Among them the canonical tensor model has a canonical pair of rank-three
tensors as dynamical variables, and is a pure constraint system 
with first-class constraints. The Poisson algebra of the first-class constraints has structure functions, and  
provides an algebraically consistent way of discretizing the Dirac first-class constraint algebra for general relativity.
This paper successfully formulates the Wheeler-DeWitt scheme of quantization of the canonical tensor model;
the ordering of operators in the constraints
is determined without ambiguity by imposing Hermiticity and covariance on the constraints,
and the commutation algebra of constraints takes essentially the same from as the classical Poisson algebra,
i.e. is first-class.
Thus one could consistently obtain, at least locally in the configuration space,
wave functions of ``universe" by solving the partial differential equations 
representing the constraints, i.e. the Wheeler-DeWitt equations for the quantum canonical tensor model. 
The unique wave function for the simplest non-trivial case is exactly and globally obtained.
Although this case is far from being realistic, the wave function has a few physically interesting features; 
it shows that  
locality is favored, and that there exists a locus of configurations with features of beginning of universe. 
\end{abstract}
\end{titlepage}

\section{Introduction}
\label{sec:introduction}
Tensor models were first proposed to describe simplicial quantum gravity in dimensions higher than two,
with the hope to extend the success of matrix models in the two-dimensional 
case \cite{Ambjorn:1990ge,Sasakura:1990fs,Godfrey:1990dt}. 
Tensor models were also extended to describe loop quantum gravity by considering group-valued
indices \cite{Boulatov:1992vp,Ooguri:1992eb,DePietri:1999bx,Freidel:2005qe,Oriti:2011jm}.\footnote{This class of tensor models are called group field theories.
For instance, see \cite{Gielen:2013kla} for a recent interesting discussion.} 
However, the original tensor models with symmetric tensors have some unfavored
features from the view point of simplicial quantum gravity \cite{Sasakura:1990fs,DePietri:2000ii}, 
and had actually been unsuccessful
in obtaining sensible analytical results for a long period of time.
This was drastically changed by the advent of colored tensor models \cite{Gurau:2009tw}, 
whose dynamical variables are unsymmetric tensors.   
The colored tensor models have good correspondence to simplicial manifolds, 
and a number of interesting analytical results have been derived so far \cite{Gurau:2011xp}.
The colored tensor models have also stimulated the developments of the renormalization procedure for
tensor group field theories \cite{BenGeloun:2011rc,BenGeloun:2012pu,Geloun:2012bz,
Carrozza:2013wda}.   
However, since the dominant simplicial manifolds in colored tensor models have been shown to be  
branched polymers \cite{Gurau:2013cbh}, 
it is remaining as a serious open question how wide-spread spaces
like our universe can be dominated in tensor models, while some interesting variants of colored tensor models
have been explored \cite{Bonzom:2012wa,Tanasa:2011ur,Dartois:2013he,Bonzom:2013lda},
and sub-dominant contributions have been studied \cite{Kaminski:2013maa}. 

A possible resolution of the problem may be obtained by considering time direction more seriously. 
The above developments of colored tensor models have basically been dealing with Euclidean signature.
While, in field theories on flat spaces, results for Minkowski signature are obtainable by analytic continuations from 
computations in Euclidean signature, it is not clear at all whether this is also true for dynamical space-time.
In fact, Causal Dynamical Triangulation (CDT), which is simplicial quantum gravity incorporating 
time direction, has numerically been shown to generate wide-spread simplicial spaces \cite{Ambjorn:2013tki},
while Dynamical Triangulation, namely the Euclidean case, is not successful in this respect.
It is also instructive that matrix models have been constructed for two-dimensional CDT \cite{Ambjorn:2008jf,Ambjorn:2012zx}.

On account of these facts suggesting the importance of time direction in quantum gravity, 
the present author has proposed a tensor model in the canonical formalism, dubbed canonical tensor 
model \cite{Sasakura:2011sq,Sasakura:2012fb,Sasakura:2013gxg}.
It has a canonical pair of rank-three 
tensors as its dynamical variables, and is defined 
as a pure first-class constraint system. 
The on-shell closure of the first-class constraint Poisson algebra guarantees
the consistency of ``many-fingered time" evolutions,
and the fact that it has structure functions makes the dynamics non-trivial.
In view of these features, the canonical tensor model is very similar to general relativity;
in the ADM formalism \cite{Arnowitt:1962hi}, general relativity is a pure constraint system with first-class constraints, 
and the Poisson algebra of these constraints has structure 
functions \cite{Dirac:1958sq,Dirac:1958sc,Arnowitt:1962hi,DeWitt:1967yk}.
In fact, this Dirac constraint algebra 
for general relativity can be obtained 
from the constraint algebra of the canonical tensor model by taking a formal locality 
limit \cite{Sasakura:2011sq}.
Since the Dirac constraint algebra plays a major role in geometrodynamics \cite{Hojman:1976vp}, 
it would be a highly interesting possibility that the canonical tensor model reproduces general relativity
in a certain, yet unknown, infrared limit.
For this to be seen, first of all, a space-like object must be dynamically generated from the canonical tensor model,
and more specifically the formal locality limit mentioned above must be derived from dynamics.
The purpose of the present paper is to take a first step in this direction
by formulating quantization of the canonical tensor model and treating the simplest non-trivial case.
 
In Section~\ref{sec:classical}, the classical canonical tensor model 
\cite{Sasakura:2011sq,Sasakura:2012fb,Sasakura:2013gxg} is reviewed.
For simplicity, 
this paper deals only with the minimal canonical tensor model, 
which was introduced in the previous paper \cite{Sasakura:2013gxg},
since the quantization is more or less the same for other non-minimal models.
In Section~\ref{sec:quantization}, the Wheeler-DeWitt scheme of quantization of the canonical tensor model is formulated.
The canonical variables are lifted to Heisenberg operators, and 
the ordering of the operators in the constraints is determined unambiguously by imposing Hermiticity and covariance 
on the constraints.
Then it is shown that the commutation algebra of the constraints is first-class.
Because of the on-shell closure of the commutation algebra, one could consistently 
obtain, at least locally in the configuration space, 
wave functions of ``universe" which satisfy the partial differential equations representing the constraints.
In Section~\ref{sec:exact}, 
the unique wave function for the simplest non-trivial case  
is exactly and globally obtained by 
solving the full set of Wheeler-DeWitt equations.
In Section~\ref{sec:interpretation}, 
the physical interpretation of the exact wave function is discussed.
The wave function shows that locality is favored and that there exists
a locus which may be interpreted as beginning of universe.  
The final section is devoted to summary and discussions.

\section{Classical canonical tensor model}
\label{sec:classical}
The dynamical variables of the minimal canonical tensor model \cite{Sasakura:2013gxg}
are a pair of real and symmetric rank-three tensors, 
$M_{abc}$ and $P_{abc}$ $(a,b,c=1,2,\ldots,N) $, 
\[
&M_{abc}=M^*_{abc}=M_{bca}=M_{cab}=M_{bac}=M_{acb}=M_{cba},\label{eq:propm}\\
&P_{abc}=P^*_{abc}=P_{bca}=P_{cab}=P_{bac}=P_{acb}=P_{cba}.
\]
They are assumed to satisfy the canonical Poisson brackets,
\[
&\{M_{abc},P_{def}\}=\sum_{\sigma} \delta_{a \,\sigma(d)}\delta_{b \,\sigma(e)}
\delta_{c \,\sigma(f)},\\
&\{M_{abc},M_{def}\}=\{P_{abc},P_{def}\}=0, 
\] 
where the summation is over all the permutations of $d,e,f$.

With these canonical variables, the constraints are expressed as 
\[ 
&{\cal H}_a=\frac{1}{2} M_{abc}M_{bde}P_{cde}, \\
&{\cal J}_{[ab]}=\frac{1}{4} \left( P_{acd}M_{bcd}-P_{bcd}M_{acd} \right ), \\
&{\cal D}=-\frac{1}{6} M_{abc} P_{abc}, \label{eq:sconst}
\]
where the repeated indices are assumed to be summed over. This convention will implicitly be used unless otherwise 
stated in this paper.
Here $[ab]$ symbolically indicates the antisymmetry, ${\cal J}_{[ab]}=-{\cal J}_{[ba]}$.
The second constraints are the infinitesimal generators of the kinematical symmetry, 
which is globally the orthogonal group $O(N)$,
\[
M'_{abc} =  L_{aa'}L_{bb'}L_{cc'} M_{a'b'c'}, \ \ L\in O(N) .
\label{eq:orthogonaltrans}
\] 
The third constraint is the generator of a scaling transformation, 
and was introduced in \cite{Sasakura:2013gxg} to prevent the typical runaway behaviors of the dynamics.
The scaling constraint makes the configuration space effectively compact, and the dynamics will behave well.
The first and second constraints may be called the Hamiltonian and momentum constraints, respectively,
expressing their corresponding physical roles in analogy with general relativity.

These constraints form a first-class constraint Poisson algebra given by
\[
&\{ H(T_1),H(T_2) \}=J([\tilde T_1,\tilde T_2]), 
\label{eq:phh}\\
&\{ J(V),H(T) \}=H(VT), 
\label{eq:pjh}\\
&\{ J(V_1),J(V_2) \}=J([V_1,V_2]), 
\label{eq:pjj}\\
&\{{\cal D}, H(T) \}=H(T), \label{eq:pdh} \\
&\{{\cal D}, J(V) \}=0, \label{eq:pdj}
\]
where
\[
&H(T)=T_a {\cal H}_a, \\
&J(V)=V_{[ab]} {\cal J}_{[ab]},
\]
with a real vector $T_a$ and an antisymmetric real matrix $V_{[ab]}=-V_{[ba]}$.
On the right-hand sides of the Poisson algebra, $\tilde T$ is a symmetric real matrix defined by 
\[
\tilde T_{bc} = T_a M_{abc},
\]
$VT$ is the multiplication of a matrix and a vector, and $[\ ,\ ]$ denotes 
the matrix commutator.  
Since the right-hand side of \eq{eq:phh} contains $\tilde T$
dependent on $M$, the algebra has structure functions, but not
structure constants. 
This feature makes the fairly simple Poisson algebra rather non-trivial and 
raises the expectation that physically interesting dynamics would occur in the canonical tensor model.

A time reversal symmetry for the model can be formulated by the invariance of the constraints 
(and therefore the constraint Poisson algebra) under
\[
&{\cal H}_a(M,P)\rightarrow -{\cal H}_a(M,- P),
\label{eq:timeh}\\
&{\cal J}_{[ab]}(M, P) \rightarrow - {\cal J}_{[ab]}( M,- P),\label{eq:timej}\\
& {\cal D}(M, P) \rightarrow - {\cal D}( M,- P) \label{eq:timed}.
\]
An important implication of this symmetry is to prevent an ambiguity in defining the scaling constraint $\cal D$.
In fact, merely for the closure of the constraint algebra,  
$\cal D$ is allowed to have an arbitrary real constant shift, ${\cal D}+d$,
but this violates the time reversal symmetry \eq{eq:timed}. 

A physical interpretation of the canonical rank-three tensor model is a Hamilton dynamical theory for fuzzy 
spaces \cite{Sasakura:2011ma,Sasakura:2005js}. 
A fuzzy space is a generalization of the concept of the so-called non-commutative space, and is defined by
an algebra of functions $f_a\ (a=1,2,\ldots, N)$ on it, the product of which is given by
\[
 f_a \star f_b=C_{ab}{}^c f_c,
 \] 
where $C_{ab}{}^c$ are the structure constants of the function algebra. Thus a fuzzy space is essentially characterized by
the rank-three tensor $C_{ab}{}^c$, and, by making it dynamical, one obtains a dynamical theory of fuzzy spaces. 
In the correspondence to the rank-three tensor model \cite{Sasakura:2011ma}, a fuzzy space is also assumed to be 
equipped with a trace-like symmetric and real inner product satisfying
\[
\langle f_a | f_b \star f_c \rangle=\langle f_a \star f_b | f_c \rangle=\langle f_c \star f_a | f_b \rangle,
\label{eq:cyclic}
\]  
and also a complex conjugation, 
\[
&f_a{}^*=f_a,  \label{eq:realityfunction}\\
&(f_a\star f_b)^*=f_b\star f_a. \label{eq:complexconjugation}
\]
Then the variable $M_{abc}$ of the rank-three tensor model can be related to the structure constants 
by identifying $M_{abc}=C_{ab}{}^d \langle f_d | f_c \rangle$ for an orthonormal basis of functions satisfying
$\langle f_a | f_b \rangle=\delta_{ab}$.
In the case of the minimal tensor model with the property \eq{eq:propm}, the function algebra is also
commutative,
\[
f_a \star f_b =f_b \star f_a. \label{eq:commutative}
\]
 
The above defining properties of fuzzy spaces are in fact naturally encountered in physical situations.
For instance, let me first consider a usual $D$-dimensional continuous space. 
On it, an arbitrary function can be expressed by a linear combination of Dirac delta functions,
\[
f^{con.}_z = \delta^D (x-z), 
\]
where $x$ and $z$ are the argument and the label of the functions, respectively, and 
they both take the values of the $D$-dimensional spatial coordinate. 
Thus the basis of functions can be given by $\{f_z^{con.}\ |\ z\in R^D\}$. 
The product of the basis functions can be computed as
\[
f^{con.}_{z_1} f^{con.}_{z_2}=\delta^D(x-z_1) \delta^D(x-z_2)=\delta^D(z_1-z_2) \delta^D(x-z_1)
= \delta^D(z_1-z_2) f^{con.}_{z_1}.
\label{eq:conspace}
\]
Therefore the function algebra of the usual continuous space has the structure constants given by Dirac delta functions. 
Also with $\langle f^{con.}_{z_1} | f^{con.}_{z_2} \rangle_{con.} =\delta^D(z_1-z_2)$, one can readily check that
the usual continuous space satisfy the above defining properties of fuzzy spaces, namely \eq{eq:cyclic},
\eq{eq:realityfunction}, \eq{eq:complexconjugation} (and also \eq{eq:commutative}).

It is also easy to give other physically natural examples with spatial ``fuzziness".
For instance, the scheme of momentum cutoff is an important regularization scheme in physics, and 
actually defines a fuzzy space.  Here the products of plane 
waves\footnote{Because of the reality condition of functions \eq{eq:realityfunction}, it is necessary to 
to consider real functions such as $e^{ipx}+e^{-ipx}$ and $i(e^{ipx}-e^{-ipx})$ rather than $e^{ipx}$. 
Here such a rather trivial and non-essential change of basis is suppressed for simplicity.}    
are chopped as
\[
e^{ipx}\star e^{iqx}=
\left\{
\begin{array}{ll}
e^{i(p+q)x}& \hbox{if } |p+q|<\Lambda, \\
0 & \hbox{otherwise},
\end{array}
\right. 
\]
where $|p|,|q| < \Lambda$.
Also with $\langle e^{ipx}|e^{iqx} \rangle_{plane}=\delta^D(p+q)$, one can readily show that all the above 
defining properties of fuzzy spaces are satisfied. 
This plane wave algebra is a commutative non-associative algebra, and encodes a geometric property
of the fuzzy space
in the algebra, because of the existence of the cutoff scale $\Lambda$. This is in sharp contrast with the 
so-called non-commutative
spaces, in which geometric properties are encoded in Dirac operators \cite{Connes:1994yd}. 

The above plane wave fuzzy space is not suited for practical computation, because of the sharp momentum cutoff. 
A more computable fuzzy space of a similar kind can be defined by a function algebra \cite{Sasai:2006ua},
\[
&f^{gauss}_{z_1} \star f^{gauss}_{z_2}= 
\int d^D z_3\ M_{z_1z_2z_3} \ f^{gauss}_{z_3}, \label{eq:ffmf}\\
&M_{z_1 z_2 z_3}=c(\beta,g) \exp\left[ -\beta\left( (z_1-z_2)^2+(z_2-z_3)^2+(z_3-z_1)^2\right)\right],
\label{eq:mgauss}
\] 
where
\[
z^2=g_{\mu\nu}\,z^\mu z^\nu
\]
with constant $g_{\mu\nu}$, and 
$c(\beta,g)$ is a coefficient dependent on $\beta$ and $g=\hbox{Det}[g_{\mu\nu}]$. 
Here the $z_i$'s are the $D$-dimensional coordinates labeling functions, 
and $\mu,\nu=1,2,\ldots,D$.  Also with $\langle f^{gauss}_{z_1} | f^{gauss}_{z_2} \rangle_{gauss} =\delta^D(z_1-z_2)$,
one can readily check that all the above defining properties of fuzzy spaces are satisfied.
The parameter $\beta$ is redundant in the sense that it can be absorbed into the redefinition of $g_{\mu\nu}$,
but is convenient in describing the scale of fuzziness by implicitly assuming $g_{\mu\nu}={\cal O}(1)$.
From the function algebra \eq{eq:ffmf} and \eq{eq:mgauss}, 
one can see that the scale of fuzziness is in the order of $1/\sqrt{\beta}$, and  
the function algebra approaches \eq{eq:conspace} of the usual continuous space in the limit $\beta \rightarrow \infty$
by appropriately choosing the normalization $c(\beta,g)$.
Thus the $\beta\rightarrow\infty$ limit can be considered to be a locality limit in which the fuzziness disappears and 
the fuzzy space becomes the usual continuous space.
Note that this notion of locality that $M_{xyz}$ dominates only at $x\sim y \sim z$ is 
dependent on the basis taken for functions and is generally 
obscured by the orthogonal transformation \eq{eq:orthogonaltrans}. 
In fact there exists an $O(N)$ invariant which measures the locality 
and it will appear in Section \ref{sec:interpretation}. 

Now let me go back to discuss the canonical tensor model.
The part \eq{eq:phh}, \eq{eq:pjh} and \eq{eq:pjj} of the Poisson algebra has some similarity to the Dirac
constraint algebra \cite{Dirac:1958sq,Dirac:1958sc,Arnowitt:1962hi,DeWitt:1967yk} for general relativity.  
In fact,  the latter can formally be derived from the 
former in the locality limit \cite{Sasakura:2011sq}. 
In the paper \cite{Sasakura:2011sq}, the initial assumption is to formally replace the tensor indices to spacial coordinates 
and consider a localized form of the tensor as given in \eq{eq:mgauss}.
By putting this form into the evaluation of the matrix commutator in the right-hand side of \eq{eq:phh}
and taking the locality limit $\beta\rightarrow \infty$ with an appropriate choice of $c(\beta,g)$,
one can reproduce the Poisson bracket between Hamiltonian constraints of the Dirac algebra of general relativity.
The other rather trivial kinematical Poisson brackets in general relativity can also be derived in similar fashions.  
Strictly speaking, the derivation in \cite{Sasakura:2011sq} is explicit merely for constant $g_{\mu\nu}$.
However, a coordinate-invariant generalization can straightforwardly be done by assuming instead a 
coordinate-invariant form \cite{Sasakura:2007ud},
\[
M_{xyz}=c(\beta) \left[g(x)g(y)g(z)\right]^\frac{1}{4}
\exp\left[ -\beta\left( d(x,y)^2+d(y,z)^2+d(z,x)^2\right)\right],
\label{eq:mgaussgen}
\]
where $g(x)=\hbox{Det}[g_{\mu\nu}(x)]$, and $d(x,y)$ is the geodesic distance between $x$ and $y$.
This generalization does not change the final result in the locality limit $\beta \rightarrow \infty$, since
the derivatives of $g_{\mu\nu}(x)$ with respect to $x$ are in higher orders 
of $1/\beta$, and vanish in the locality limit.  
Moreover the Gaussian form of \eq{eq:mgauss} is not essential; it can be any form which 
respects a kind of locality, $M_{xyz}$ dominates only in the neighborhood of 
$x\sim y\sim z$, with a strict locality limit $\beta\rightarrow \infty$. 
Certainly the derivation depends on such formal assumptions, and it is necessary to have a dynamical reason for
them.
This is an important purpose of studying the dynamics of the canonical tensor model.
It will be shown in Section~\ref{sec:interpretation} that the locality is indeed favored dynamically in the simplest 
non-trivial case.

What looks peculiar in \eq{eq:timeh} is the minus sign in front of ${\cal H}_a$ in the right-hand side. 
This comes from the fact that ${\cal H}_a$ is linear in $P$.
On the other hand, the Hamiltonian constraint in general relativity 
does not have the minus sign under the time reversal transformation,
since it is quadratic in $\pi^{\mu\nu}(x)$,
which is the momentum conjugate to the metric tensor field $g_{\mu\nu}(x)$.
So, to delete the minus sign in \eq{eq:timeh},
one would then tend to propose $M\rightarrow -M$ instead of  $P\rightarrow -P$
under the time reversal transformation.
Though this possibility cannot absolutely be denied, it seems 
implausible, because the assumption \eq{eq:mgaussgen} used in the derivation 
of the Dirac algebra in general relativity does not seem compatible with the time reversal 
transformation of $g_{\mu\nu}(x)$. 
This qualitative difference between ${\cal H}_a$ and the Hamiltonian constraint
in general relativity suggests that, if existed,  
the relation between them would be more involved than what
superficially looks in the forms of the Poisson algebras. 
Even if so, there still exists good motivation for studying the dynamics of the canonical tensor model
in relation to general relativity,
since the dynamics of the canonical tensor model in the formal locality limit and that of the general relativity are 
controlled by the same Dirac or the hypersurface deformation algebra of the general relativity
\cite{Arnowitt:1962hi,Dirac:1958sq,Dirac:1958sc,DeWitt:1967yk,Hojman:1976vp}.

Finally, it would be worth giving comments on the rank of tensors. 
The rank-three is intimately related to the interpretation that the tensor model is a dynamical theory of fuzzy spaces,
as explained above. Since fuzzy spaces can in principle approximate 
any dimensional spaces, the rank-three
tensor model is expected to contain any dimensional quantum gravity.
Therefore, though a rank-three tensor is the minimum and simplest tensor over a matrix, the rank-three tensor model
should be enough for the purpose of quantum gravity. On the other hand, there is a good motivation
to consider higher rank tensors in view of simplicial quantum gravity. Presently, however, it seems hard to 
proceed for the higher rank tensors in the same manner as the rank-three tensor model. The construction and 
the proof of uniqueness
of the hamiltonian constraints for the rank-three tensor model have been carried out by brute force analysis in 
\cite{Sasakura:2012fb,Sasakura:2011sq}. 
This kind of analysis becomes much more complicated and difficult for higher rank tensors; 
the closure condition of the constraint algebra generally 
becomes much more difficult to be satisfied, because much more terms are generated 
from computations of Poisson brackets.  
In fact, a simple analogical trial such as ${\cal H}_a=M_{abcde} M_{bcghi} P_{deghi}$ for 
rank-five symmetric tensors does not seem to work.
One would need a more sophisticated and systematic methodology for the analysis.   

\section{Quantization}
\label{sec:quantization}
There exist two main quantization schemes for gravity, 
the reduce phase space quantization and the Wheeler-DeWitt one.
As discussed in \cite{Thiemann:2004wk}, the former scheme 
would be inappropriate in incorporating fluctuating time. 
Thus this paper takes the Wheeler-DeWitt scheme for quantization.

In quantization, the canonical variables are lifted to Heisenberg operators as 
\[
&[\hM_{abc},\hP_{def}]=i \sum_{\sigma} \delta_{a \,\sigma(d)}\delta_{b \,\sigma(e)}
\delta_{c \,\sigma(f)},
\label{eq:comhmhp}\\
&[\hM_{abc},\hM_{def}]=[\hP_{abc},\hP_{def}]=0.
\]
The operators are assumed to satisfy the properties corresponding to the classical case,
\[
&\hM_{abc}=\hM^\dagger_{abc}=\hM_{bca}=\hM_{cab}=\hM_{bac}=\hM_{acb}=\hM_{cba},
\label{eq:prophm}\\
&\hP_{abc}=\hP^\dagger_{abc}=\hP_{bca}=\hP_{cab}=\hP_{bac}=\hP_{acb}=\hP_{cba},
\]
where $^\dagger$ denotes Hermitian conjugate of operators.

The ordering of operators in the constraints can be determined as follows.
As for the momentum and scaling constraints, by imposing Hermiticity, one uniquely obtains
\[
\hJ_{[ab]}&=\frac{1}{8} \left( \hP_{acd}\hM_{bcd}+\hM_{bcd}\hP_{acd}-\hP_{bcd}\hM_{acd} -\hM_{acd}\hP_{bcd} \right ) 
\CR
&=\frac{1}{4}\left(\hM_{bcd}\hP_{acd}-\hM_{acd}\hP_{bcd}\right), \\
\hD&=-\frac{1}{12} \left(\hM_{abc} \hP_{abc}+\hP_{abc}\hM_{abc} \right) \CR
&=-\frac{1}{6}\left(\hM_{abc} \hP_{abc}-i \frac{N(N+1)(N+2)}{2}\right),
\]
because of their quadratic forms.
Here this paper takes the convention that $\hP$ is placed in the rightmost. 
This is because $\hP$ will explicitly be represented by partial derivatives $\partial/\partial M$ 
in the following section.

The ordering of operators in the Hamiltonian constraint is a bit more involved
because of its cubic form. However, 
the covariance with respect to the kinematical symmetry, or the consistency with
the momentum constraint $\hat {\cal J}$, requires that it must have the form,
\[
\hH_a=\frac{1}{2}\left(\hM_{abc} \hM_{bde} \hP_{cde} - i \lambda_{\cal H} \hM_{abb}\right),
\label{eq:formhH}
\]
where $\lambda_{\cal H}$ is a real parameter to be 
determined.
Then, by imposing Hermiticity on $\hH$, it is determined that
\[
\lambda_{\cal H}=\frac{(N+2)(N+3)}{2}.
\]

An important question is whether these quantized constraints are mutually consistent.
This can be checked by computing the commutation algebra among these constraints. 
Since $\hJ$ and $\hD$ merely generate linear Lie transformations on $\hM$ and $\hP$, the commutators containing
$\hJ$ or $\hD$ are rather trivial, i.e. take the same forms as the classical Poisson algebra.
Thus the only non-trivial commutator is between $\hH$.
However, in this case too, because the first term of $\hH$ in \eq{eq:formhH} is linear in $\hP$, 
one readily realizes that the commutator between the first terms in \eq{eq:formhH}
takes the same form as the classical Poisson
algebra, if the convention that $\hP$'s are placed in the rightmost is kept. So the only non-trivial 
computation is the mixing between the first and second terms in \eq{eq:formhH}, which actually vanishes 
as
\[
[\hM_{abc}\hM_{bde}\hP_{cde},\hM_{a'b'b'}]-(a \leftrightarrow a')&=4 \hM_{abc} \hM_{a'bc}+2 \hM_{aa'b}\hM_{bcc}
-(a\leftrightarrow a') =0.
\]
Thus the whole commutation algebra basically takes the same form as the classical Poisson algebra,
and is explicitly given by
\[
&[ \hqH(T_1),\hqH(T_2) ]=i \hqJ([{\hat{T}_1},{\hat{T}_2}]), 
\label{eq:qhh}\\
&[\hqJ(V),\hqH(T)]=i \hqH(VT), 
\label{eq:qjh}\\
&[\hqJ(V_1),\hqJ(V_2) ]=i \hqJ([V_1,V_2]), 
\label{eq:qjj}\\
&[\hD, \hqH(T)]=i\hqH(T), \label{eq:qdh} \\
&[\hD, \hqJ(V)]=0, \label{eq:qdj}
\]
where 
\[
&\hqH(T)=T_a \hH_a, \\
&\hqJ(V)=V_{[ab]} \hJ_{[ab]}, \\ 
&\hqJ(\hat V)={\hat V}_{[ab]} \hJ_{[ab]}, \label{eq:hqj}\\
&\hat T_{bc} = T_a \hM_{abc}.
\]
Here one has to follow the ordering of the operators, which are indicated with $\hat{\ }$, as written above.
Thus the quantized constraints form a first-class commutation algebra,
and the Wheeler-DeWitt wave function $\Psi$ could consistently be obtained by
solving the Wheeler-DeWitt equations,
\[
\hH_a \Psi=\hJ_{[ab]} \Psi=\hD \Psi=0,
\label{eq:wdeq}
\]
in an appropriate representation of operators.

The classical time reversal symmetry \eq{eq:timeh}, \eq{eq:timej} and \eq{eq:timed} can be extended to a
quantum version,
 \[
&\hat {\cal H}_a(\hat M,\hat P)\rightarrow -\hat{\cal H}^*_a(\hat M,-\hat P),\\
&\hat{\cal J}_{[ab]}(\hat M,\hat P) \rightarrow -\hat{\cal J}^*_{[ab]}(\hat M,-\hat P),\\
&\hat {\cal D}(\hat M,\hat P) \rightarrow -\hat {\cal D}^*(\hat M,-\hat P),
\]
where $^*$ denotes complex conjugation.

\section{The exact wave function for $N=2$}
\label{sec:exact}
This section solves the Wheeler-DeWitt equations \eq{eq:wdeq} explicitly for 
the simplest non-trivial case. Consider a representation of the operators in terms of $M$ as
\[
&\Psi=\Psi(M), \\
&\hM_{abc}=M_{abc},\\
&\hP_{abc}=-i \Delta(abc)\frac{\partial}{\partial M_{abc}},
\]
where $\Delta(abc)$ is a multiplicity factor defined by
\[
\Delta(abc)=
\left\{
\begin{aligned}
6 & \hbox{ for } a=b=c,\\
2 & \hbox{ for } a=b\neq c,b=c\neq a,c=a\neq b,\\
1 & \hbox{ for } a\neq b, b\neq c, c\neq a.
\end{aligned}
\right.
\]
This factor is needed to consistently take account of the properties \eq{eq:comhmhp} and \eq{eq:prophm}.
Then the Wheeler-DeWitt equations \eq{eq:wdeq} are a set of first-order partial differential equations
on $\Psi(M)$.

The total number of the first-order partial differential equations in \eq{eq:wdeq} is given by $N+N(N-1)/2+1=(N^2+N+2)/2$,
while the number of degrees of freedom of $M_{abc}$ satisfying \eq{eq:propm} is $N(N+1)(N+2)/6$. 
At $N=2$ they both take the same number 4 , and the Wheeler-DeWitt equations will have 
a unique solution, granted that it is also globally consistent.
Below it will explicitly be obtained. 

An efficient way to explicitly solve the equations is to first solve them on a certain subspace 
in the configuration space and then extend it to the whole space. 
This is equivalent to what is usually called gauge-fixing, which, in this case, is supposed to be done
for $\hD$ and $\hJ$. 
As such a subspace, consider
\[
\label{eq:sub1}
&M_{111}=1,\\
\label{eq:sub2}
&M_{112}=0,\\
\label{eq:sub3}
&M_{122}=x_1,\\
\label{eq:sub4}
&M_{222}=x_2,
\]
where $x_1$ and $x_2$ are real.
On the subspace, the explicit expressions of the Wheeler-DeWitt equations \eq{eq:wdeq} are given by
\[
&\left [
\frac{\partial}{\partial M_{111}}+x_1 \frac{\partial}{\partial x_1}+x_2 \frac{\partial}{\partial x_2}+2\right] \Psi=0, \\
&\left [
(1-2x_1)\frac{\partial}{\partial M_{112}}-x_2 \frac{\partial}{\partial x_1}+3 x_1 \frac{\partial}{\partial x_2}\right] \Psi=0, \\
&\left [
3\frac{\partial}{\partial M_{111}}+ x_1(1+2x_1) \frac{\partial}{\partial x_1}+3 x_1 x_2 \frac{\partial}{\partial x_2}+5(1+x_1)\right] \Psi=0, \\
&\left [
x_1(1+2 x_1)\frac{\partial}{\partial M_{112}}+  3 x_1 x_2 \frac{\partial}{\partial x_1}+3 ({x_1}^2+ {x_2}^2) \frac{\partial}{\partial x_2}+5 x_2 \right] \Psi=0, 
\]
where the first and the second equations come from $\hD \Psi=0$ and $\hJ_{12} \Psi=0$, respectively, 
and the last two ones from $\hH_a \Psi=0$. 
By using the first and second equations to solve for $\partial \Psi/\partial M_{111}$ and $\partial \Psi/ \partial M_{112}$
\footnote{This corresponds to that variables conjugate to gauge-fixing conditions will
also be fixed for first-class constraints.},
and putting them into the last two equations, one obtains
\[
&\left [2 x_1 ( x_1-1) \frac{\partial}{\partial x_1}+3 x_2 (x_1-1) \frac{\partial}{\partial x_2}+5 x_1-1\right] \Psi=0, \label{eq:x1x2eq1}\\
&\left [4 x_1 x_2 ( x_1-1) \frac{\partial}{\partial x_1}+3 (4 {x_1}^3+2 x_1 {x_2}^2-{x_2}^2) \frac{\partial}{\partial x_2}+5 x_2 (2 x_1-1)\right] \Psi=0.
\label{eq:x1x2eq2}
\]

The two equations \eq{eq:x1x2eq1} and \eq{eq:x1x2eq2} can be combined to delete the non-derivative terms and obtain
\[
\left [ x_1 x_2  (x_1-1) \frac{\partial}{\partial x_1}+2 (5 {x_1}^4-{x_1}^3+2 x_1 {x_2}^2 -{x_2}^2) \frac{\partial}{\partial x_2}\right] \Psi=0.
\]
This implies that $\Psi$ is constant along the trajectories satisfying
\[
\frac{dx_2}{dx_1}=\frac{2 (5 {x_1}^4-{x_1}^3+2 x_1 {x_2}^2 -{x_2}^2)}{ x_1 x_2  (x_1-1)},
\]
which can elementarily be solved by noticing that it can be deformed to
\[
\frac{d }{d x_1}{x_2}^2=\frac{4(2x_1-1)}{x_1(x_1-1)} {x_2}^2+\frac{4{x_1}^2(5 x_1-1)}{x_1-1}.
\]
The solution is 
\[
a_0=\frac{4 {x_1}^3+{x_2}^2}{{x_1}^4 (x_1-1)^4},
\label{eq:a0x1x2}
\]
with a constant $a_0$. This implies that $\Psi$ depends only on the specific combination of $x_1$ and $x_2$ indicated in \eq{eq:a0x1x2}
as 
\[
\Psi=f\left(\frac{4 {x_1}^3+{x_2}^2}{{x_1}^4 (x_1-1)^4}\right)
\]
with a function $f$. Putting this expression back to \eq{eq:x1x2eq1}, $f$ is determined to be $f(x) \propto \sqrt{x}$, and 
finally the wave function on the subspace defined by \eq{eq:sub1} through \eq{eq:sub4} is obtained as 
\[
\Psi=b_0\frac{\sqrt{4 {x_1}^3+{x_2}^2}}{{x_1}^2 (x_1-1)^2}
\label{eq:exact}
\]
with a constant $b_0$\footnote{Strictly speaking, since the first-derivatives of $\Psi$ are 
divergent at $4 {x_1}^3+{x_2}^2=0$, it is not clear whether the constant factors $b_0$ in both the regions divided by
$4 {x_1}^3+{x_2}^2=0$ must be identical or not. This ambiguity may be deleted by an analytic continuation with respect to
$x_1,x_2$ out of real values.} .
Extension of \eq{eq:exact} to the whole configuration space will be discussed in the following section. 

\section{Physical interpretation of the exact wave function}
\label{sec:interpretation}
The number of degrees of freedom for $N=2$ is so small that this case is 
very far from being realistic.
Nonetheless it will be seen that the wave function
shows interesting features concerning locality and beginning of universe. 
The expression of the wave function in the whole configuration space will also be obtained in this section.

Since the wave function \eq{eq:exact} 
has infinite peaks where the denominator vanishes, there exists a kind of preference for such configurations. 
To find their physical meaning, let me first rewrite the denominator in the form invariant under the kinematical symmetry $O(N)$.
One can explicitly check that the denominator can be rewritten as 
\[
x_1{}^2(x_1-1)^2&=\frac{1}{2} \left(
M_{acd}M_{bde}M_{bef}M_{afc}
-M_{acd}M_{bde}M_{aef}M_{bfc}
\right)
\label{eq:expmmmm}
 \\
&=-\frac{1}{4}\sum_{a,b} \hbox{Tr}\left([M^{(a)},M^{(b)}]^2\right) 
\label{eq:commm}
\]
on the subspace defined by \eq{eq:sub1} through \eq{eq:sub4}, 
where $M^{(a)}$ are real symmetric matrices defined by
\[
M^{(a)}{}_{bc}=M_{abc}.
\]
The expression \eq{eq:expmmmm} can be used in the whole configuration space out of the subspace.

Because of the semi-positive definiteness of the expression \eq{eq:commm} for general $M$, its vanishing is equivalent to
the mutual commutativity of the matrices,
\[
[M^{(a)},M^{(b)}]=0,
\]
for all $a,b$.
This implies that the real symmetric matrices $M^{(a)}$ can simultaneously be 
diagonalized by an orthogonal group transformation of the kinematical symmetry.
Then, by also taking into account \eq{eq:propm}, 
such $M_{abc}$ can be shown to be transformed to a symmetric diagonal form,
\[
M_{abc}=m_a \delta_{ab} \delta_{ac},
\label{eq:diagonal}
\]   
with real $m_a$ by the orthogonal group transformation.

As reviewed in Section~\ref{sec:classical}, 
the canonical tensor model can be interpreted to describe Hamiltonian dynamics of 
fuzzy spaces \cite{Sasakura:2011ma,Sasakura:2005js}. 
In this interpretation, 
$M_{abc}$ correspond to the structure constants defining a fuzzy space characterized by an algebra,
\[
f_a \star f_b=M_{abc}f_c,
\] 
among functions $f_a\ (a=1,2,\cdots,N)$ on a fuzzy space.
Physically speaking, functions can be regarded as fuzzy ``points" in an appropriate basis of functions, 
and in the case of the diagonal form \eq{eq:diagonal},
the ``points" are mutually independent as
\[
f_a \star f_b =m_a \delta_{ab} f_a.
\]
Thus the expression \eq{eq:commm} can be considered to be a measure of locality among ``points" forming 
a fuzzy space.
Then the divergent configurations of the wave function \eq{eq:exact} correspond to the fuzzy spaces in which locality is maximized.

Next, I will discuss the vanishing locus of the wave function \eq{eq:exact}. 
The wave function gets multiplied by the imaginary unit, when $x_1,x_2$ passes through the vanishing locus,
$4 {x_1}^3+{x_2}^2=0$.
This behavior of the wave function has some similarity to what are often discussed in literatures \cite{Grishchuk:1993ds}
on beginning of universe born from nothing, since there exists a sort of discontinuity in the wave function. 
A more mathematically definite characteristic of the vanishing locus can be given as follows.
One can explicitly check that, at $4 {x_1}^3+{x_2}^2=0$, 
there exists a real vector $v_a$ such that the determinant of the symmetric matrix $v_a M^{(a)}$ vanishes.
In this sense, the configuration $M$ is degenerate at the locus,
which would agree with the interpretation that the locus is the beginning of ``universe",
since our universe should have started from a point-like state (or a very small state).
One can readily write the above statement in an $O(N)$-invariant fashion to reproduce the
numerator of \eq{eq:exact} as
\[
-\hbox{Det}[\tilde M]
&=-\frac{1}{2} \epsilon_{ac}\epsilon_{bd} \epsilon_{eg} \epsilon_{fh} \epsilon_{e'g'} \epsilon_{f'h'} M_{aef}M_{bgh} M_{ce'f'}M_{dg'h'}
\label{eq:numexp} \\
&=4 x_1{}^3+x_2{}^2
\]
on the subspace, 
where $\epsilon_{12}=-\epsilon_{21}=1,\epsilon_{11}=\epsilon_{22}=0$, and
\[
\tilde M_{ab}=\epsilon_{ce} \epsilon_{df} M_{acd} M_{bef}.
\]

By collecting \eq{eq:expmmmm} and \eq{eq:numexp}, 
the expression of the wave function valid in the whole configuration space is given by
\[
\label{eq:phiwhole}
\Psi=c_0\, 
\frac{
\sqrt{\epsilon_{ac}\epsilon_{bd} \epsilon_{eg} \epsilon_{fh} \epsilon_{e'g'} \epsilon_{f'h'} M_{aef}M_{bgh} M_{ce'f'}M_{dg'h'}
}}
{M_{acd}M_{bde}M_{bef}M_{afc}
-M_{acd}M_{bde}M_{aef}M_{bfc}}
\]
with a numerical constant $c_0$.  
One can explicitly check that \eq{eq:phiwhole} indeed satisfies all the un-gauged 
Wheeler-DeWitt equations \eq{eq:wdeq}.  

\section{Summary and discussions}
\label{sec:summary}
In this paper, the Wheeler-DeWitt scheme of quantization 
of the minimal canonical tensor model has successfully been formulated.
The classical constraints have been lifted to quantum ones, and their commutation algebra
has been shown to 
basically take the same form as the classical one, i.e.~be first-class.
These constraints form a consistent set of Wheeler-DeWitt equations, and the wave function of 
``universe" can be obtained by solving them.     
Indeed the unique wave function for $N=2$ has explicitly been obtained and 
its physical interpretation has been discussed.
Although the case of $N=2$ is far from being realistic,
the wave function shows physically interesting features 
such as that locality is favored, and that there exists a locus of configurations
which have characteristics of beginning of universe. 

An obvious future question is whether the physically interesting properties of the wave function
found in $N=2$ can be generalized to $N\geq 3$ or not. 
Especially, 
as is reviewed in Section~\ref{sec:classical}, the emergence of locality is essentially important for 
the constraint algebra of the canonical tensor model to be identified with Dirac or the hypersurface deformation algebra
of general relativity. 
Since the numbers of the dynamical variables and the constraints are in the order of ${\cal O}(N^3)$ and ${\cal O}(N^2)$†,
respectively, there will exist larger freedom for wave functions for larger $N$, if possible global conditions 
are assumed to be ignorable. This freedom may potentially diminish the possibility that
the rather clear physical properties found for $N=2$ also appear for large $N$. 
Therefore it would be important to find qualitative understanding of the properties of the wave function in $N=2$
and to see whether it can be generalized to any $N$ or not. 
In addition, a new aspect in $N\geq 3$ is that
the configuration space contains points where part of the kinematical symmetry is unbroken,
while, for $N=2$, the $O(2)$ symmetry is broken all over the whole configuration space.
Since symmetry is a key feature in our physical understanding of nature, it would be highly interesting
to study fates of symmetry, i.e. whether symmetry is favored or not and how, in the canonical tensor model.
Such studies are also expected to connect the canonical tensor model to tensor group field 
theories \cite{Boulatov:1992vp,Ooguri:1992eb,DePietri:1999bx,Freidel:2005qe,Oriti:2011jm,Gielen:2013kla,BenGeloun:2011rc,BenGeloun:2012pu,Geloun:2012bz,Carrozza:2013wda},
in which specific groups are chosen as inputs.

The framework of the canonical tensor model is very similar to that of general relativity, and 
there exists ``problem of time" in the quantization formulated in this paper.
Thus it should be interesting to apply the
resolutions developed for general relativity, such as complete 
observables \cite{Rovelli:2001bz,Rovelli:1990ph,Dittrich:2005kc,Dittrich:2004cb,Tambornino:2011vg},  
to the canonical tensor model, and see how they work.    
An advantage of the canonical tensor model in this respect is that it
is a finite system and therefore can be computed in principle without any divergences.
An important question to be answered would be how the standard unitary evolution 
can be recovered by appropriate choices of time and observables.

Concerning the above question on time, 
what seems a problem in the exact wave function for $N=2$ is that there exist no oscillations. 
In our daily life, time is measured by counting oscillations, and,
also from theoretical view points,  it seems effective to formulate time and distance by oscillations.
Moreover, oscillations are widely observed in nature, and if a natural theory did not contain them, it should be 
discarded. 
At this stage it is unclear whether oscillations can be found in $N\geq 3$ or not.
However, since there will exist larger freedom of wave functions for larger $N$,
one will probably be able to consider oscillatory boundary conditions of wave functions
for larger $N$. 
It would be interesting to see whether oscillatory solutions to the Wheeler-DeWitt equations 
actually exist or not for $N \geq 3$ under such boundary conditions.

The above issue of oscillatory solutions seems also related to the question how the Hamiltonian constraint 
of general relativity may be generated from the canonical tensor model,
as briefly raised in Section~\ref{sec:classical}. 
Oscillatory solutions are more often found in 
the second order differential equations rather than in the first order ones as the Wheeler-DeWitt equations
of the canonical tensor model.   
Then a possibility is that these oscillatory solutions would be described 
by effective second order differential equations,
which may be identified with the Wheeler-DeWitt equations of general relativity.
This may not be as fancy as it sounds, since there exists a good real example;
the Klein-Gordon equation, which is second order, is derivable from the Dirac equation, which is first order.
In this, an essence is neglecting the spin degrees of freedom.
Since the canonical tensor model obviously contains much more degrees of freedom than what
are necessary for describing field theories, 
it could be related to general relativity, only after 
integrating over its huge part of degrees of freedom. 
The existence of the constraint algebra in the canonical tensor model, which is indeed comparable to 
Dirac or the hypersurface deformation algebra of general relativity,
would play essential roles in such possibility.



\begin{thebibliography}{99}
\bibitem{Ambjorn:1990ge}
  J.~Ambjorn, B.~Durhuus and T.~Jonsson,
  ``Three-Dimensional Simplicial Quantum Gravity And Generalized Matrix
  Models,''
  Mod.\ Phys.\ Lett.\ A {\bf 6}, 1133 (1991).

\bibitem{Sasakura:1990fs}
  N.~Sasakura,
  ``Tensor Model For Gravity And Orientability Of Manifold,''
  Mod.\ Phys.\ Lett.\ A {\bf 6}, 2613 (1991).

\bibitem{Godfrey:1990dt}
  N.~Godfrey and M.~Gross,
  ``Simplicial Quantum Gravity In More Than Two-Dimensions,''
  Phys.\ Rev.\ D {\bf 43}, 1749 (1991).
  
\bibitem{Boulatov:1992vp}
  D.~V.~Boulatov,
  ``A Model of three-dimensional lattice gravity,''
  Mod.\ Phys.\ Lett.\ A {\bf 7}, 1629 (1992)
  [arXiv:hep-th/9202074].

\bibitem{Ooguri:1992eb}
  H.~Ooguri,
  ``Topological lattice models in four-dimensions,''
  Mod.\ Phys.\ Lett.\ A {\bf 7}, 2799 (1992)
  [arXiv:hep-th/9205090].

\bibitem{DePietri:1999bx}
  R.~De Pietri, L.~Freidel, K.~Krasnov and C.~Rovelli,
  ``Barrett-Crane model from a Boulatov-Ooguri field theory over a  homogeneous
  space,''
  Nucl.\ Phys.\ B {\bf 574}, 785 (2000)
  [arXiv:hep-th/9907154].
  
\bibitem{Freidel:2005qe} 
  L.~Freidel,
  ``Group field theory: An Overview,''  Int.\ J.\ Theor.\ Phys.\  {\bf 44}, 1769 (2005)  [hep-th/0505016].  

\bibitem{Oriti:2011jm} 
  D.~Oriti,
  ``The microscopic dynamics of quantum space as a group field theory,''  arXiv:1110.5606 [hep-th].  
 
\bibitem{Gielen:2013kla} 
  S.~Gielen, D.~Oriti and L.~Sindoni,
  ``Cosmology from Group Field Theory,''
  arXiv:1303.3576 [gr-qc].

\bibitem{DePietri:2000ii} 
  R.~De Pietri and C.~Petronio,
  ``Feynman diagrams of generalized matrix models and the associated manifolds in dimension 4,''
  J.\ Math.\ Phys.\  {\bf 41}, 6671 (2000)
  [gr-qc/0004045].
 
\bibitem{Gurau:2009tw} 
  R.~Gurau,
  ``Colored Group Field Theory,''  Commun.\ Math.\ Phys.\  {\bf 304}, 69 (2011)  
  [arXiv:0907.2582 [hep-th]].  

\bibitem{Gurau:2011xp} 
  R.~Gurau and J.~P.~Ryan,
  ``Colored Tensor Models - a review,''
  SIGMA {\bf 8}, 020 (2012)
  [arXiv:1109.4812 [hep-th]].

\bibitem{BenGeloun:2011rc} 
  J.~Ben Geloun and V.~Rivasseau,
  ``A Renormalizable 4-Dimensional Tensor Field Theory,''
  Commun.\ Math.\ Phys.\  {\bf 318}, 69 (2013)
  [arXiv:1111.4997 [hep-th]].
  
\bibitem{BenGeloun:2012pu} 
  J.~Ben Geloun and D.~O.~Samary,
  ``3D Tensor Field Theory: Renormalization and One-loop $\beta$-functions,''
  arXiv:1201.0176 [hep-th].

\bibitem{Geloun:2012bz} 
  J.~B.~Geloun and E.~R.~Livine,
  ``Some classes of renormalizable tensor models,''
  arXiv:1207.0416 [hep-th].
  
\bibitem{Carrozza:2013wda} 
  S.~Carrozza, D.~Oriti and V.~Rivasseau,
  ``Renormalization of an SU(2) Tensorial Group Field Theory in Three Dimensions,''
  arXiv:1303.6772 [hep-th].
 
\bibitem{Gurau:2013cbh} 
  R.~Gurau and J.~P.~Ryan,
  ``Melons are branched polymers,''
  arXiv:1302.4386 [math-ph].
    
\bibitem{Bonzom:2012wa} 
  V.~Bonzom,
  ``New 1/N expansions in random tensor models,''
  arXiv:1211.1657 [hep-th].
 
\bibitem{Tanasa:2011ur} 
  A.~Tanasa,
  ``Multi-orientable Group Field Theory,''
  J.\ Phys.\ A {\bf 45}, 165401 (2012)
  [arXiv:1109.0694 [math.CO]].
 
\bibitem{Dartois:2013he} 
  S.~Dartois, V.~Rivasseau and A.~Tanasa,
  ``The 1/N expansion of multi-orientable random tensor models,''
  arXiv:1301.1535 [hep-th].
  
\bibitem{Bonzom:2013lda} 
  V.~Bonzom and F.~Combes,
  ``Fully packed loops on random surfaces and the 1/N expansion of tensor models,''
  arXiv:1304.4152 [hep-th].
  
\bibitem{Kaminski:2013maa} 
  W.~Kaminski, D.~Oriti and J.~P.~Ryan,
  ``Towards a double-scaling limit for tensor models: probing sub-dominant orders,''
  arXiv:1304.6934 [hep-th].
    
\bibitem{Ambjorn:2013tki} 
  J.~Ambjorn, A.~Goerlich, J.~Jurkiewicz and R.~Loll,
  ``Quantum Gravity via Causal Dynamical Triangulations,''
  arXiv:1302.2173 [hep-th].
 
\bibitem{Ambjorn:2008jf} 
  J.~Ambjorn, R.~Loll, Y.~Watabiki, W.~Westra and S.~Zohren,
  ``A Matrix Model for 2D Quantum Gravity defined by Causal Dynamical Triangulations,''
  Phys.\ Lett.\ B {\bf 665}, 252 (2008)
  [arXiv:0804.0252 [hep-th]].
 
\bibitem{Ambjorn:2012zx} 
  J.~Ambjorn, L.~Glaser, A.~Gorlich and Y.~Sato,
  ``New multicritical matrix models and multicritical 2d CDT,''
  Phys.\ Lett.\ B {\bf 712}, 109 (2012)
  [arXiv:1202.4435 [hep-th]].
  
\bibitem{Sasakura:2011sq} 
  N.~Sasakura,
  ``Canonical tensor models with local time,''
  Int.\ J.\ Mod.\ Phys.\ A {\bf 27}, 1250020 (2012)
  [arXiv:1111.2790 [hep-th]].
  
\bibitem{Sasakura:2012fb} 
  N.~Sasakura,
  ``Uniqueness of canonical tensor model with local time,''
  Int.\ J.\ Mod.\ Phys.\ A {\bf 27}, 1250096 (2012)
  [arXiv:1203.0421 [hep-th]].
  
\bibitem{Sasakura:2013gxg} 
  N.~Sasakura,
  ``A canonical rank-three tensor model with a scaling constraint,''
    Int.\ J.\ Mod.\ Phys.\ A {\bf 28}, 1350030  (2013)
  [arXiv:1302.1656 [hep-th]].
 
\bibitem{Arnowitt:1962hi} 
  R.~L.~Arnowitt, S.~Deser and C.~W.~Misner,
  ``The Dynamics of general relativity,''  gr-qc/0405109.  

\bibitem{Dirac:1958sq} 
  P.~A.~M.~Dirac,
  ``Generalized Hamiltonian dynamics,''
  Proc.\ Roy.\ Soc.\ Lond.\ A {\bf 246}, 326 (1958).

\bibitem{Dirac:1958sc} 
  P.~A.~M.~Dirac,
  ``The Theory of gravitation in Hamiltonian form,''
  Proc.\ Roy.\ Soc.\ Lond.\ A {\bf 246}, 333 (1958).
  
\bibitem{DeWitt:1967yk} 
  B.~S.~DeWitt,
  ``Quantum Theory of Gravity. 1. The Canonical Theory,''  Phys.\ Rev.\  {\bf 160}, 1113 (1967).  

\bibitem{Hojman:1976vp} 
  S.~A.~Hojman, K.~Kuchar and C.~Teitelboim,
  ``Geometrodynamics Regained,''  Annals Phys.\  {\bf 96}, 88 (1976).  

\bibitem{Connes:1994yd} 
  A.~Connes,
  ``Noncommutative geometry,''
  ISBN-9780121858605.

\bibitem{Sasai:2006ua} 
  Y.~Sasai and N.~Sasakura,
  ``One-loop unitarity of scalar field theories on Poincare invariant commutative nonassociative spacetimes,''
  JHEP {\bf 0609}, 046 (2006)
  [hep-th/0604194].

\bibitem{Sasakura:2007ud} 
  N.~Sasakura,
  ``The Lowest modes around Gaussian solutions of tensor models and the general relativity,''
  Int.\ J.\ Mod.\ Phys.\ A {\bf 23}, 3863 (2008)
  [arXiv:0710.0696 [hep-th]].
 
\bibitem{Thiemann:2004wk} 
  T.~Thiemann,
  ``Reduced phase space quantization and Dirac observables,''
  Class.\ Quant.\ Grav.\  {\bf 23}, 1163 (2006)
  [gr-qc/0411031].
  
\bibitem{Sasakura:2011ma} 
  N.~Sasakura,
  ``Tensor models and 3-ary algebras,''
  J.\ Math.\ Phys.\  {\bf 52}, 103510 (2011)
  [arXiv:1104.1463 [hep-th]].

\bibitem{Sasakura:2005js} 
  N.~Sasakura,
  ``An Invariant approach to dynamical fuzzy spaces with a three-index variable,''
  Mod.\ Phys.\ Lett.\ A {\bf 21}, 1017 (2006)
  [hep-th/0506192].
  
\bibitem{Grishchuk:1993ds} 
  L.~P.~Grishchuk,
  ``Quantum effects in cosmology,''
  Class.\ Quant.\ Grav.\  {\bf 10}, 2449 (1993)
  [gr-qc/9302036].
  
\bibitem{Rovelli:2001bz} 
  C.~Rovelli,
  ``Partial observables,''
  Phys.\ Rev.\ D {\bf 65}, 124013 (2002)
  [gr-qc/0110035].
  
\bibitem{Rovelli:1990ph} 
  C.~Rovelli,
  ``What Is Observable In Classical And Quantum Gravity?,''
  Class.\ Quant.\ Grav.\  {\bf 8}, 297 (1991).
  
\bibitem{Dittrich:2005kc} 
  B.~Dittrich,
  ``Partial and complete observables for canonical general relativity,''
  Class.\ Quant.\ Grav.\  {\bf 23}, 6155 (2006)
  [gr-qc/0507106].
  
\bibitem{Dittrich:2004cb} 
  B.~Dittrich,
  ``Partial and complete observables for Hamiltonian constrained systems,''
  Gen.\ Rel.\ Grav.\  {\bf 39}, 1891 (2007)
  [gr-qc/0411013].
  
\bibitem{Tambornino:2011vg} 
  J.~Tambornino,
  ``Relational Observables in Gravity: a Review,''
  SIGMA {\bf 8}, 017 (2012)
  [arXiv:1109.0740 [gr-qc]].
  
  \end{thebibliography}
  \end{document}